\documentclass[journal]{IEEEtran}
\usepackage{times}
\usepackage{url}
\usepackage[noadjust]{cite}
\usepackage[dvips]{graphicx}
\usepackage{psfrag}
\usepackage{subfig}
\usepackage{amsmath}
\usepackage{amsthm}
\usepackage[font=footnotesize,labelfont=bf]{caption}
\usepackage{algorithmic,algorithm,indentfirst}
\usepackage{units}
\usepackage{color,soul}
\usepackage{epstopdf}

\begin{document}

\title{Substitutability of Spectrum and Cloud-based Antennas in Virtualised Wireless Networks}

\author{\IEEEauthorblockN{Hamed Ahmadi\IEEEauthorrefmark{1}, Irene Macaluso\IEEEauthorrefmark{2}, Ismael Gomez\IEEEauthorrefmark{2}, Linda Doyle\IEEEauthorrefmark{2}, Luiz DaSilva \IEEEauthorrefmark{2}}\\
\IEEEauthorblockA{\IEEEauthorrefmark{1}{School of Electrical and Electronic Engineering, University College Dublin, Ireland}\\ \IEEEauthorrefmark{2}{CONNECT, the Center for Future Networks, Trinity College Dublin, Ireland}} \\

\thanks{This material is based upon works supported by
the Science Foundation Ireland under Grants No. 10/CE/I1853 and 10/IN.1/I3007, and by the Seventh Framework Programme for Research of the European Commission under grant number ADEL-619647.}}
\maketitle

\begin{abstract}

Some of the new trends emerging in future wireless networks enable a vastly increased fluidity in accessing  a wide range of resources, thus supporting flexible network composition and dynamic allocation of resources to virtual network operators (VNOs).
In this work we study a new resource allocation opportunity that is enabled by the cloud radio access network architecture. In particular, we investigate the relationship between the cloud-based antennas and spectrum as two important resources in virtualized wireless networks. We analyze the interplay between spectrum and antennas in the context of an auction-based allocation mechanism through which VNOs can bid for a combination of the two types of resources. Our analysis shows that the complementarity and partial substitutability of the two resources significantly impact the results of the allocation of those resources and uncovers the possibility of divergent interests between the spectrum and the infrastructure providers.

\end{abstract}

\IEEEpeerreviewmaketitle

\section{Introduction}

Extreme virtualization \cite{forde2011}, sees future networks as entities composed out of consumable commodities, many of which will be shared and where the boundary between pure software and hardware resources is blurred. This pool of commodities consists of highly heterogeneous resources: infrastructure, remote radio heads (RRHs), spectrum, storage, processing power, information content, knowledge, backhaul, hypervisors, etc. In our vision, future virtual network operators (VNOs) construct their networks from this shared pool of resources, which enables them to dynamically use and release resources according to their demand \cite{doyle2014NwoB}.


%
In this work, we investigate the complex relationships between network resources and a mechanism to allocate resources to VNOs. In particular, we study the substitutability of spectrum and massive cloud-based antennas. These two types network resources are partially substitutable, which means that spectrum and cloud-based antennas are interchangeable only after the VNO has acquired a minimum required amount of spectrum and number of antennas. The partial substitutability of these resources and the dynamic demand of VNOs for them introduce a new and interesting resource management problem which is different from the conventional methods of spectrum and/or infrastructure management.
%


We consider the cloud-based massive-MIMO (Multiple Input Multiple Output) antennas as a resource that multiple VNOs can share at the same time. The shared infrastructure could be provided by the existing network operators to reduce their operational costs. Deploying a public network is another option that can contribute to the pool of shared infrastructure \cite{ALUEcoShare}. Dynamic infrastructure sharing also creates opportunities for new technology ventures and opens the market to  small size and local infrastructure providers.

On the other hand, in order to allow operators to maintain some quality assurances in deploying new services, we assume that spectrum resources should be orthogonally allocated to the operators. Spectrum can be made available through the new frameworks that are envisioned in the 3.5 GHz Citizens Broadband Radio Service  \cite{fcc2015} and Licensed Shared Access (LSA) licensing scheme. This approach of spectrum and infrastructure management is in line with the radio access network (RAN) sharing scenario proposed by the 3rd Generation Partnership Project (3GPP) in \cite{3gppTr22.951}, where the RAN is shared between the participating operators while each of them transmits on orthogonal portions of the licensed spectrum. 

\begin{figure*}
	\centering
	\includegraphics[width=1\textwidth]{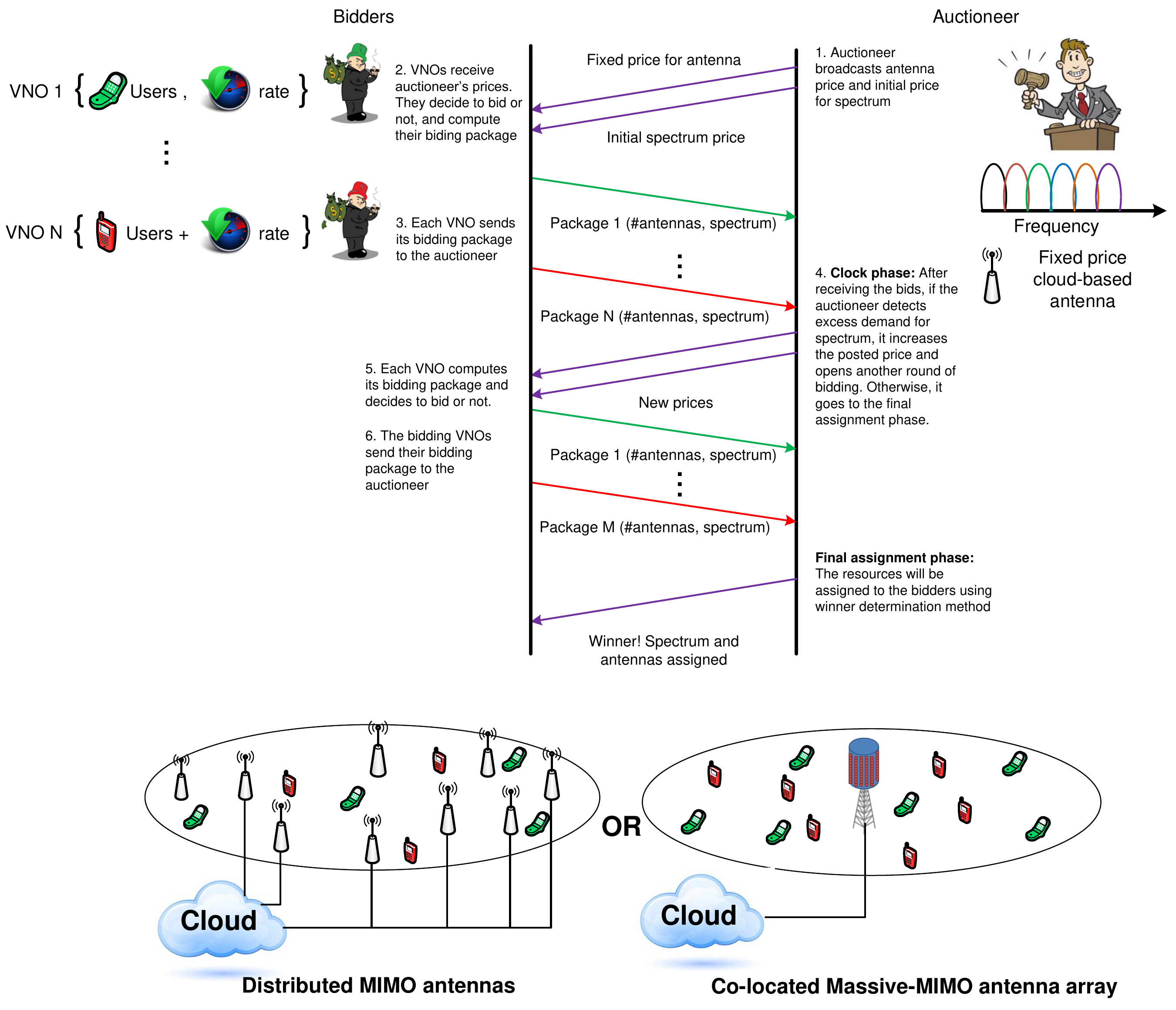}
	\caption{The proposed ecosystem; VNOs request antennas (at a fixed price) and bid for spectrum to satisfy the rate required by their subscribers. The cloud-based antennas can be collocated massive MIMO antennas or fiber-connected distributed MIMO antennas.}\label{fig:sysmodel}
\end{figure*}

In our approach, spectrum is auctioned to the demanding VNOs and its final price is determined by the market. In particular, we propose a hybrid auction model which is performed by a third party auctioneer for the combined acquisition of spectrum and antennas. The results reveal a complex interplay between antennas and spectrum in an auction-based allocation, and show the presence of potential divergent interests between the spectrum and the infrastructure providers. The auction mechanism and the possible scenarios of having collocated and distributed cloud-based massive-MIMO antennas are summarised in Figure~\ref{fig:sysmodel}. We will discuss these scenarios and the auction mechanism in the following sections. We begin with a brief introduction to market-based resource allocation in wireless networks in Section~\ref{sec:lit}. Major features of the Cloud RAN architecture are described in Section~\ref{sec:cran}. Section~\ref{sec:auction} describes the auction-based allocation mechanism through which VNOs can bid for a combination of antennas and spectrum resources. Section~\ref{sec:results} analyzes the impact of the complex relation between the two types of resources on an auction-based allocation mechanism. We summarize our conclusions and point towards directions for future work in Section~\ref{sec:conc}.

\section{Literature review} \label{sec:lit}

Auctioning is a well-known approach for selling goods to the highest bidder, and in the context of resource management it is an effective way of allocating the resources to the ones who value them the most \cite{zhang2013auction}. Unlike auction-based approaches, in non-market-based models normally a central entity optimizes its utility subject to some constraints on the minimum data rate required by the VNOs \cite{van2014dynamic}. Generally such centralized systems treat requests from the VNOs similarly and satisfy a fairness metric. Therefore, in this type of systems the resource assignment decision is made centrally, without direct participation from VNOs, while in the market-based approaches the VNOs decide their bids and the resources are assigned to the VNOs that value them the most. Moreover, auction-based approaches typically have a lower computational complexity than the centralized approaches, since bids are decided in a distributed fashion.

Previous research on market-based resource allocation in wireless networks has focused on homogeneous radio resources, typically owned by a single entity \cite{zhang2013auction}. For example, auction models have been proposed to allocate downlink bandwidth provided by a single network operator \cite{dramitinos2010auction}. Auction theory has also been used to model access to secondary spectrum \cite{zhou2009trust}. Other works have focused on auctioning network capacity provided by one network operator \cite{doyleWhitepaper}. Only recently the auctioning of multiple heterogeneous resources has started to attract attention \cite{Zhu_CombAuction15}. The authors of that work proposed a hierarchical Vickrey-Clarke-Groves auction to jointly assign infrastructure and spectrum and analyzed the allocation efficiency of the proposed schemes. In our work, similarly to \cite{Zhu_CombAuction15}, we auction both antennas and spectrum, but with two important differences. First and foremost, the focus in this paper is on the effects of the complementarity and partial substitutability of spectrum and antennas on the auction-based resource allocation scheme, with special attention to the implications of disjoint ownership of those resources.  Secondly, in our model we consider the cloud-based massive-MIMO antennas as a resource that multiple VNOs can share at the same time. This is particularly important in view of the RAN sharing scenarios proposed by 3GPP in \cite{3gppTr22.852}.

\section{CRAN Architecture} \label{sec:cran}

The concept of extreme sharing intertwines with the design of submissive network components: components which are not biased towards any particular technology or model of usage, but can be configured dynamically to accommodate a range of functionalities. In light of this, we have chosen to focus on the Cloud RAN (CRAN) architecture as the basis of our discussion.

In a CRAN, baseband processing units are decoupled from the radio frontends and located in a remote facility. This allows for a centralized and scalable operation and management of resources, including processing resources, antennas, radio frontends, etc. Software radio and cloud computing technologies are combined with centralization to fully exploit the scalability and flexibility of the network architecture. 

Most other works on CRAN consider a pool of baseband processing resources only. In this paper we consider a CRAN architecture where a switch interconnects (a subset of) the processing units in the cloud to (a subset of) the RRHs. Together with software radio and cloud computing principles, this architecture allows any processing unit to dynamically select the antennas to transmit signals to (or receive signals from). 

The potential for advanced resource management a CRAN can offer contrasts with what can be done in a classical non-CRAN architecture. In traditional networks, the baseband processing units are co-located with the antenna in the base station and designed to operate in a fixed portion of the spectrum using a fixed number of antennas, therefore supporting a limited number of users. As a consequence, network capabilities are delimited during the planning or network deployment phase.

The CRAN is a multi-antenna system where antennas can either be co-located or distributed (virtual antenna array). In both cases, users can be spatially-multiplexed if the number of users is less than or equal the number of antennas \cite{scaling_up_mimo}. In practice, the multiplexing gain is limited by the spatial correlation between channel vectors associated to different users and by the ability of the network to acquire channel state information. 

A useful property of multi-antenna systems is that the average user data rate increases with the number of antennas, even if the user is equipped with a single antenna. The rate is known to increase logarithmically in co-located antenna systems as long as user spatial signatures are still orthogonal. The behaviour in distributed systems remains unknown for the general case. For the particular case where antenna selection and regularized zero forcing precoding are employed, it is shown in \cite{joint_power_and_antenna_selection} that the achievable rate increases logarithmically with the number of antennas. Therefore, in the remaining of the paper we will assume this logarithmic relation between rate and number of antennas, which is valid for distributed or co-located systems. 

Antennas and spectrum become fungible resources capable to interchangeably provide a service to the user (rate). For example, consider a user equipped with a single-antenna terminal requesting a 1 Mbps service. Assuming a signal-to-noise (SNR) ratio of 10 dB, the network needs to designate approximately 3 MHz to that user if only 1 antenna is used. Alternatively, the network can use 10 antennas to jointly transmit the signal to that user and needs to designate only 1.5 MHz, with the additional advantage of reusing the same channel to transmit to 9 other users. In this study, we will ignore the multiplexing gains and focus on the additional gains obtained by an excess number of antennas (power gains). A similar tradeoff is possible with transmission power: the same rate can be achieved by an excess of transmission power or frequency or spatial degrees of freedom. For simplicity, throughout this paper we will assume constant power allocation for all users. 

When looking at many users being served by a VNO, the problem is essentially the same: the VNO needs to provide a sum-rate to its $K$ users, which is achieved by allocating a certain amount of spectrum and number of antennas. While the users of the same VNO share the same spectrum, the spectrum cannot be shared among VNOs. The main limitation here is the fact that each VNO has channel knowledge about its users and computes the precoding matrix based on that. Subscribers to other VNOs are seen as interference and thus we assume each VNO uses a different portion of the spectrum. This limitation does not occur with the antennas. Different signals can easily be transmitted at different frequencies using the same antenna; thus, VNOs can share the antennas whereas they cannot share the spectrum.

\section{The Auction Process}\label{sec:auction}


The auction process described here and illustrated by Figure~\ref{fig:sysmodel} is a hybrid model for the combined acquisition of spectrum and antennas. In our model antennas can be shared among all the operators, while spectrum resources are orthogonally assigned to each operator. There is a fee associated to the usage of each antenna per operator, which affects the spectrum cost. This is because spectrum and antennas are partially substitute goods and, at the same time, complementary goods. In fact, as discussed in the previous section, in a CRAN architecture spectrum and antennas are partially interchangeable resources. This implies that the demand for one resource will vary with the cost associated to the other resource. Antennas and spectrum are also complementary goods. For example, this means that if the cost incurred for the acquisition of infrastructure is too high, the remaining budget might not be sufficient to acquire the spectrum resources necessary to deliver a given service.  In view of the interconnection between these two commodities, we designed a hybrid auction model that allows operators to bid for a combination of antennas and spectrum. 

We have chosen a clock auction approach for the simplicity it affords to the bidder in putting winning packages together and for the way that it allows bidders to quickly discover prices in a complex market. The  clock auction operates in two main phases, the price discovery (clock) phase and the final assignment phase, described in the remainder of the section.

In the clock phase the auctioneer repeatedly posts a fixed price for a unit of spectrum ($1$ KHz) on sale to the bidders in a series of rounds, while the antenna price is kept constant throughout the auction. The term clock is used as the price is monotonically ascending, i.e. it can only be ticked upwards during the clock phase of the auction. In each round the bidders can indicate that they will buy packages of the items at the prices indicated. In our model, each bidder determines the number of antennas and the amount of spectrum that satisfy a minimum rate requirement, while minimizing the cost. The cost is a linear combination of the selected number of antennas and bandwidth at the prices indicated by the auctioneer. If this cost is lower than the  buyer's private valuation,  the buyer submits its package bid to the auctioneer. If the auctioneer detects excess demand for the spectrum after a round of bidding has closed, it increases the posted price and opens another round of bidding. Again, the bidders can indicate that they will buy packages of the items at the higher price indicated. As the price of spectrum increases at each round, each bidder asks for less spectrum and more antennas, due to the antenna/spectrum partial interchangeability in a CRAN. During the clock phase the bidders discover prices at near the competitive equilibrium, i.e. prices that roughly match supply with demand. In fact the clock price discovery phase approximates a second-price auction. We have modeled our implementation of the clock price discovery phase in our auction on that proposed in \cite{porter2003combinatorial}.

The clock phase of the auction ends when all excess demand is removed from the market; in an ideal situation the prices would stop rising when supply exactly equals demand. However, in complex multi-item-unit, multi-item-type auctions it is unlikely that this will occur. Hence, the approach used in this part of the auction may result in the oversupply of spectrum at the final spectrum price. This will happen if the bidders' private valuation of the minimum required rate is lower than the corresponding cost to acquire spectrum and antennas at the demanded clock price. If this situation arises then the bids are assigned using a revenue maximizing approach, i.e. using a winner determination algorithm to determine which combination of the bids that stood at the last clock price which caused excess demand will maximise the auctioneer's revenue.

The winner determination problem is formulated using the branch-on-bids approach. Bids are chosen so that the combined ``antenna+spectrum'' revenue is maximized and the constraint on the available spectrum is satisfied. The branching factor of the search tree is two and its depth is at most equal to the number of bids. Since this formulation is exponential in the number of bids, some instances of the winner determination problem may require too long to get the optimal solution. However, if an anytime search algorithm is adopted, a feasible solution can be obtained by terminating the algorithm after a certain amount of time. At each round each bidder puts together a package of antennas and spectrum that satisfy the bidder's rate requirement whilst minimizing the cost. This is an integer optimization problem. Since the rate is monotonically increasing with the number of antennas and spectrum, the rate constraint is always satisfied with equality and the problem can be converted into the minimization of a monotonic function of one variable (number of antennas), with complexity linear with the number of antennas.

\section{Numerical Results}\label{sec:results}


In this section we put the described concepts into action and discuss the auction-based mechanism described in Section~\ref{sec:auction}. We simulate a scenario with $20$ VNOs, each one  bidding to acquire a package of spectrum and infrastructure resources necessary to deliver a given service to their users, which are randomly distributed in the considered area. There is an infrastructure provider which owns $64$ antennas. In the presented simulations, we consider the $M=64$ antennas are co-located in a massive multi-antenna array. Under the assumptions discussed in Section~\ref{sec:cran}, similar results are obtained if the $64$ single-antenna RRH are randomly distributed in the same area.
%
Pilot contamination effects are ignored and backhaul is assumed ideal. The CRAN employs a TDD scheme with pilot-based training, with full channel state information (CSI) being known at the CRAN and the number of users $K$ satisfies $M\geq K$.
The total available spectrum is $50$ MHz. In the remainder of this section, the requirements for a service are expressed in terms of minimum rate requirement. Each operator assigns a constant maximum utility to the acquisition of $1$ Kbps. This means that each operator's overall budget is proportional to the required rate. All operators are homogeneous in terms of budget, minimum rate requirement, and subscribers distribution.

\begin{figure}
\centering
\includegraphics[width=0.9\columnwidth]{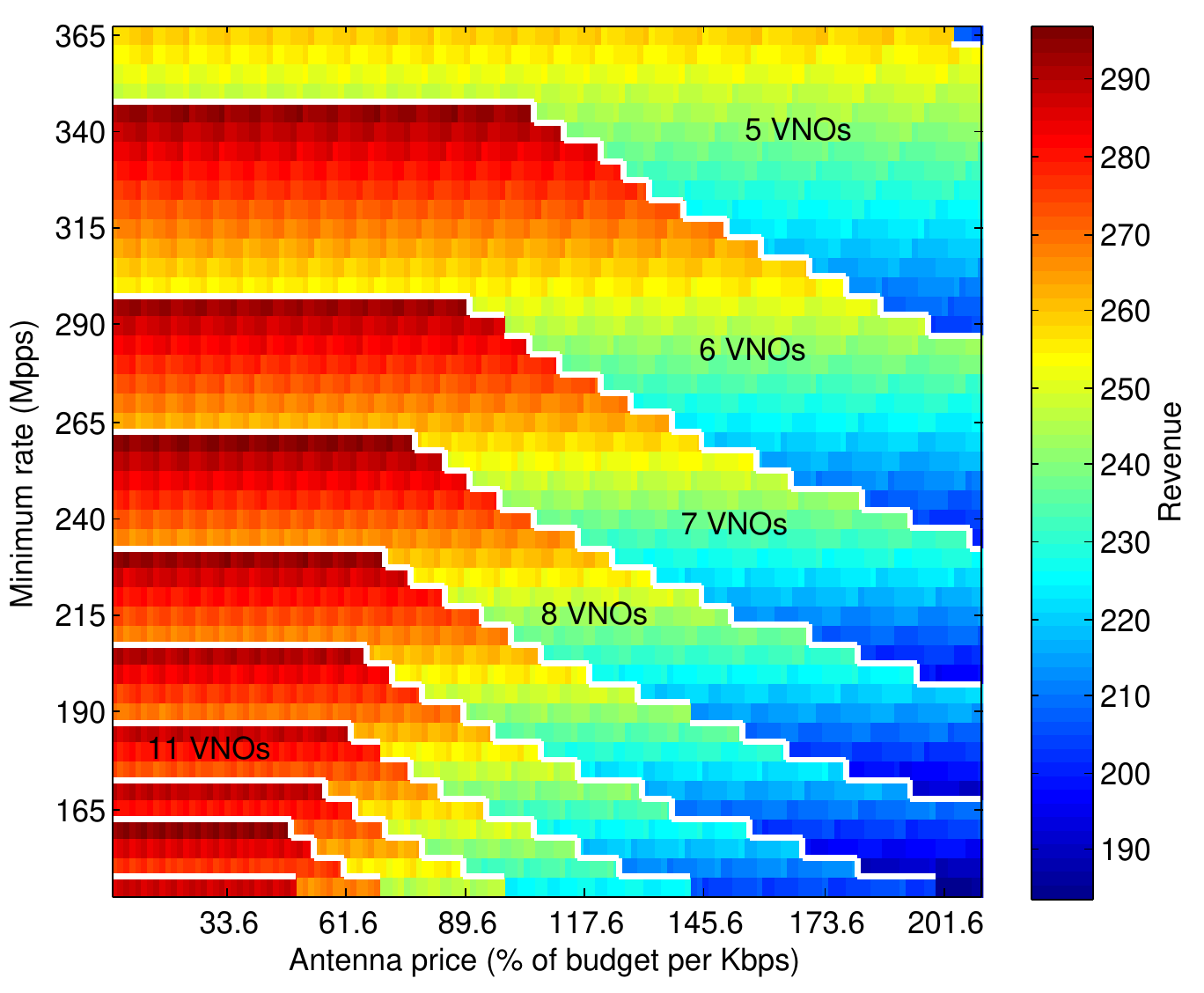}
\caption{\footnotesize Combined "antenna+spectrum" revenue and the number of VNOs that are able to obtain resources. Each pixel color shows the
combined revenue corresponding to a particular rate requirement (y axis) and antenna cost (x axis). Regions corresponding to
different numbers of winning VNOs are also highlighted.}\label{fig:SpectRev}
\end{figure}

 In the remainder of this section, we study the effects of different antenna prices and minimum required rate on a number of system performance metrics. We start by analysing the overall revenue and the number of allocated VNOs, both shown in Figure~\ref{fig:SpectRev}, as a function of the minimum required rate and antenna cost (expressed as a percentage of each VNO's budget per Kbps). As expected, the number of winning VNOs decreases when the minimum required rate increases, for a given antenna cost. In fact, by increasing their demands, the number of bids that can be accommodated by the capacity of the pool of combined resources decreases. However, the number of VNOs that successfully bid for resources also decreases when the cost of antennas increases, for a given minimum rate. This trend is not due to the physical limit of the  auctioned resources. We observe this behavior because high antenna prices cause the VNOs to demand more spectrum; since the spectrum is limited and not shared, the spectrum requirement of all VNOs will not be satisfied. We can see a clear dependence of the revenue, shown by colors in Figure~\ref{fig:SpectRev}, on  the number of winning VNOs, for any given minimum requested rate. For example, let us  consider a minimum rate of $340$ Mbps. We observe that the revenue is constant until the antenna price reaches the value of the $113\%$ of each Kbps budget. At this point the number of allocated VNOs drops (from $6$ to $5$) and so does the revenue. It is also worth noting that, inside each region identified  by the number of winning bids, the revenue increases with the requested rate, for any given antenna cost.

\begin{figure}
\centering
\includegraphics[width=0.9\columnwidth]{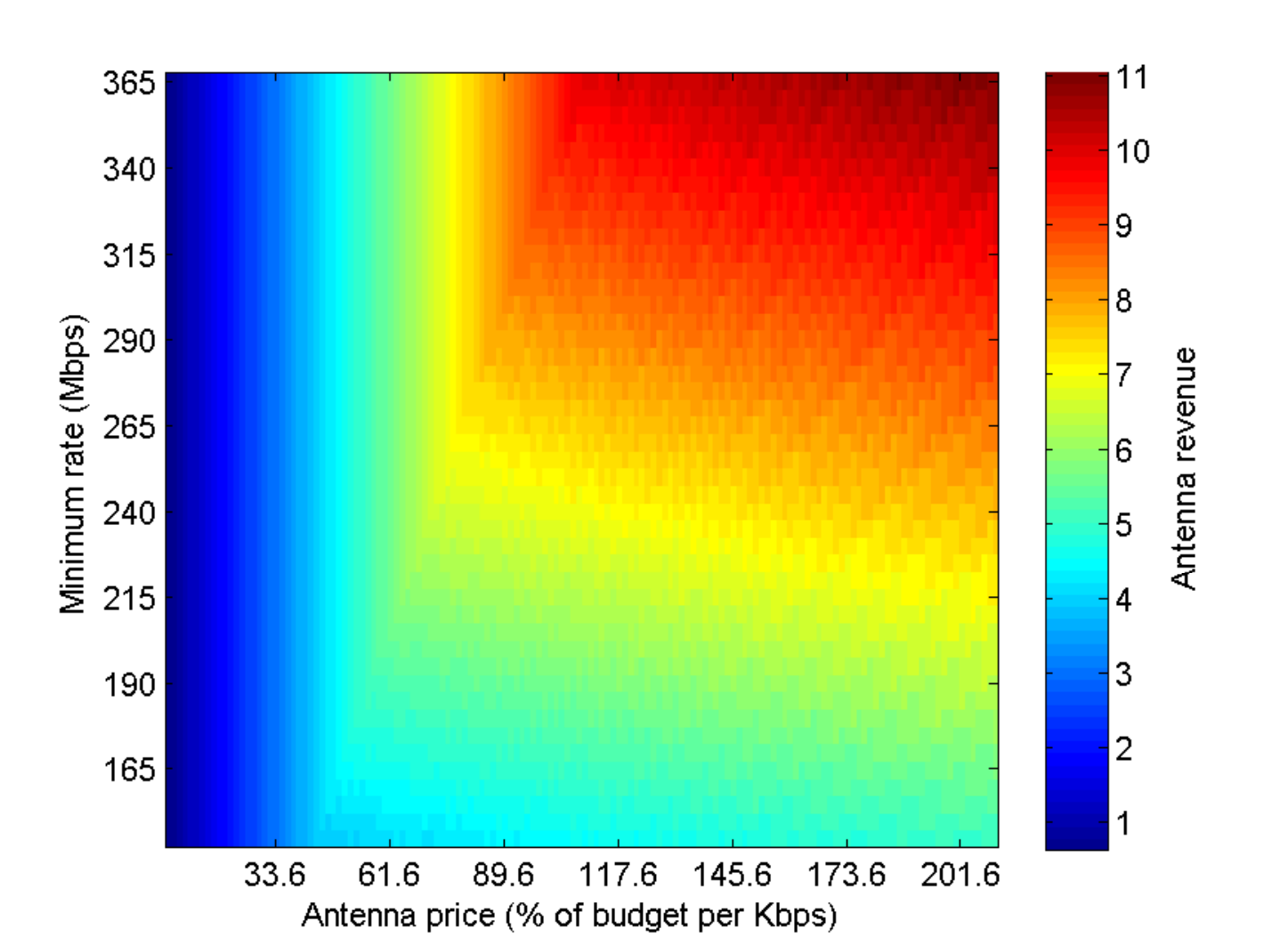}
\caption{\footnotesize Revenue resulting from antennas. Each pixel color shows the antenna revenue corresponding to a
particular rate requirement (y axis) and antenna cost (x axis).}\label{fig:AntennaRev}
\end{figure}

From the analysis of the results shown in Figure~\ref{fig:SpectRev}, it would seem clear that a lower antenna cost benefits both the auctioneer - as it results in higher revenue independently of the requested rate - and the bidders - as it increases the number of winning bids. However, the revenue shown in Figure~\ref{fig:SpectRev} is a combined revenue resulting from the combined auction of two types of resources. In the case of multiple ownership of the two resources, i.e. when the infrastructure and spectrum providers are distinct entities, it is interesting to look at the distribution of the revenue for the two types of resources. Figure~\ref{fig:AntennaRev} shows the revenue received from the allocation of antennas. For any given minimum required rate, the antenna revenue increases with the antenna cost. By comparing Figures~\ref{fig:SpectRev} and \ref{fig:AntennaRev} we should note that the combined revenue and the antenna revenue show opposing trends. This implies the presence of divergent interests between the spectrum and the infrastructure provider.

\begin{figure}
\centering
\includegraphics[width=0.9\columnwidth]{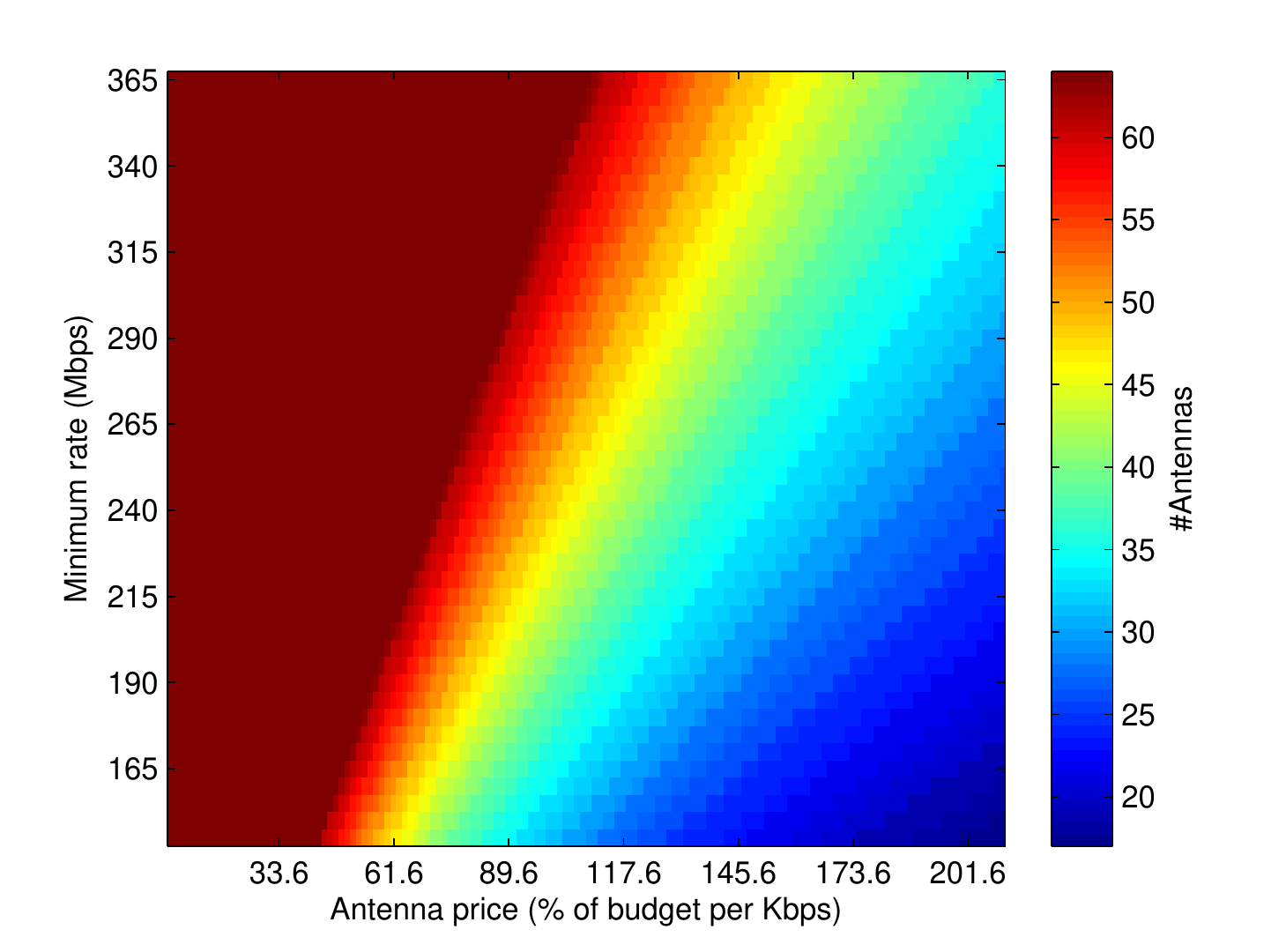}
\caption{\footnotesize Number of antennas assigned to each allocated VNO in correspondence to a particular rate requirement
(y axis) and antenna cost (x axis).}\label{fig:MeanAntenna}
\end{figure}

\begin{figure}
\centering
\includegraphics[width=0.9\columnwidth]{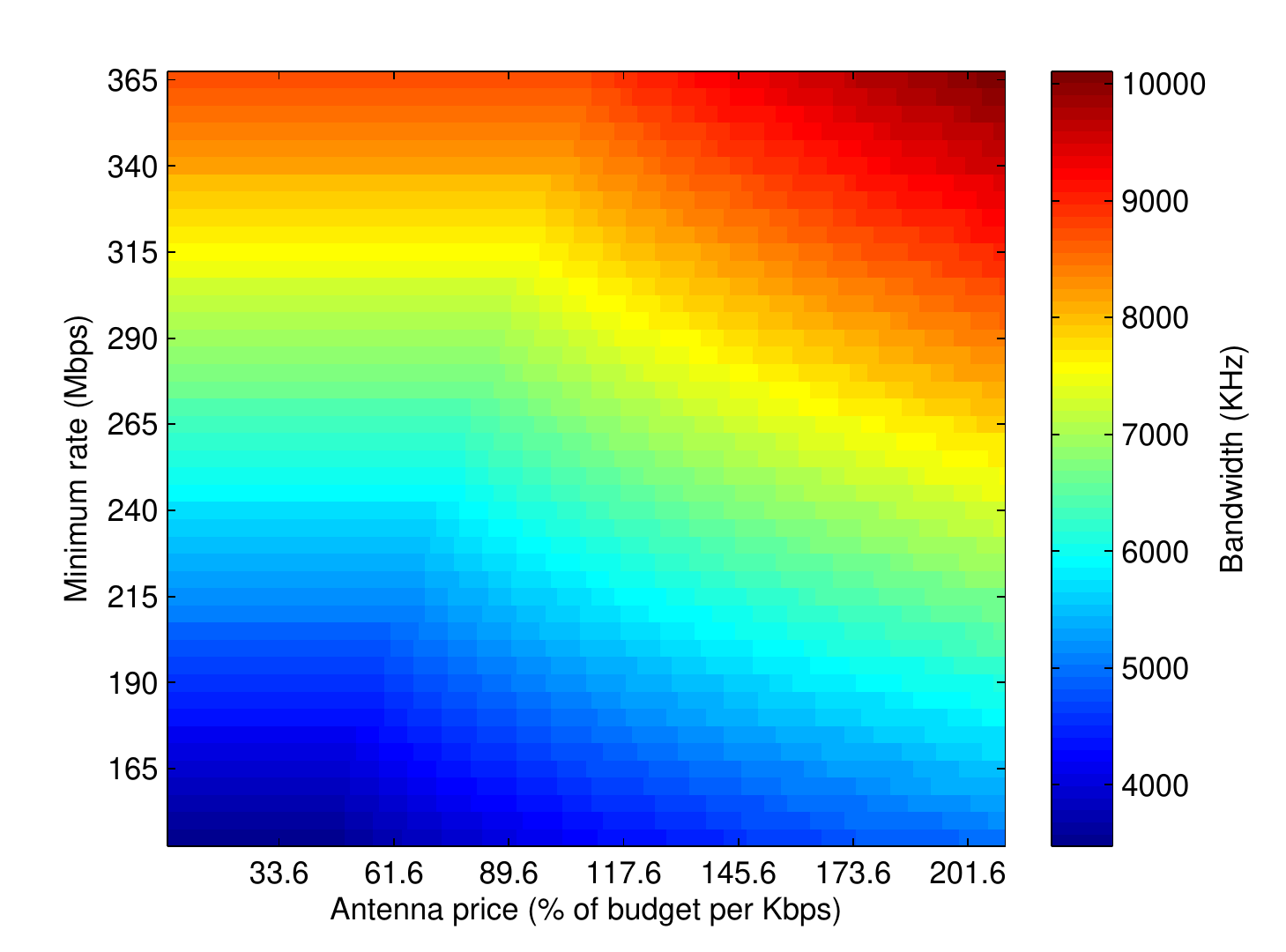}
\caption{\footnotesize Bandwidth assigned to each allocated VNO in correspondence to a particular rate requirement (y axis)
and antenna cost (x axis).}\label{fig:MeanSpectrum}
\end{figure}

We now look into the effects of our auction mechanism on the resources allocated to each VNO, namely antennas and bandwidth, respectively shown in Figures~\ref{fig:MeanAntenna} and \ref{fig:MeanSpectrum}. These results highlight the partial interchangeability of the two types of resources under consideration and its implications for the combined assignment of those resources. When the antenna cost is low, for a range of cost values depending on the rate requirement, each winning VNO is allocated the maximum number of available antennas; after a certain antenna cost is reached, the\textit{ number of allocated antennas} is constantly \textit{decreasing}, for any given rate requirement. The \textit{bandwidth} requested by each VNO for any given rate requirement is at first constant (in correspondence to the same range of values for which the maximum number of antennas is allocated) and then \textit{increases} with the antenna cost. As expected, both allocated antennas and spectrum decrease with the minimum rate, for a given antenna cost. However, in the case of antennas, this only happens after a certain antenna cost is reached. This implies that when antennas are cheap enough, in case of decrease in the rate requirement it is more convenient for the bidders to acquire less bandwidth whilst still bidding for the maximum number of antennas.

\begin{figure}
\centering
\includegraphics[width=0.9\columnwidth]{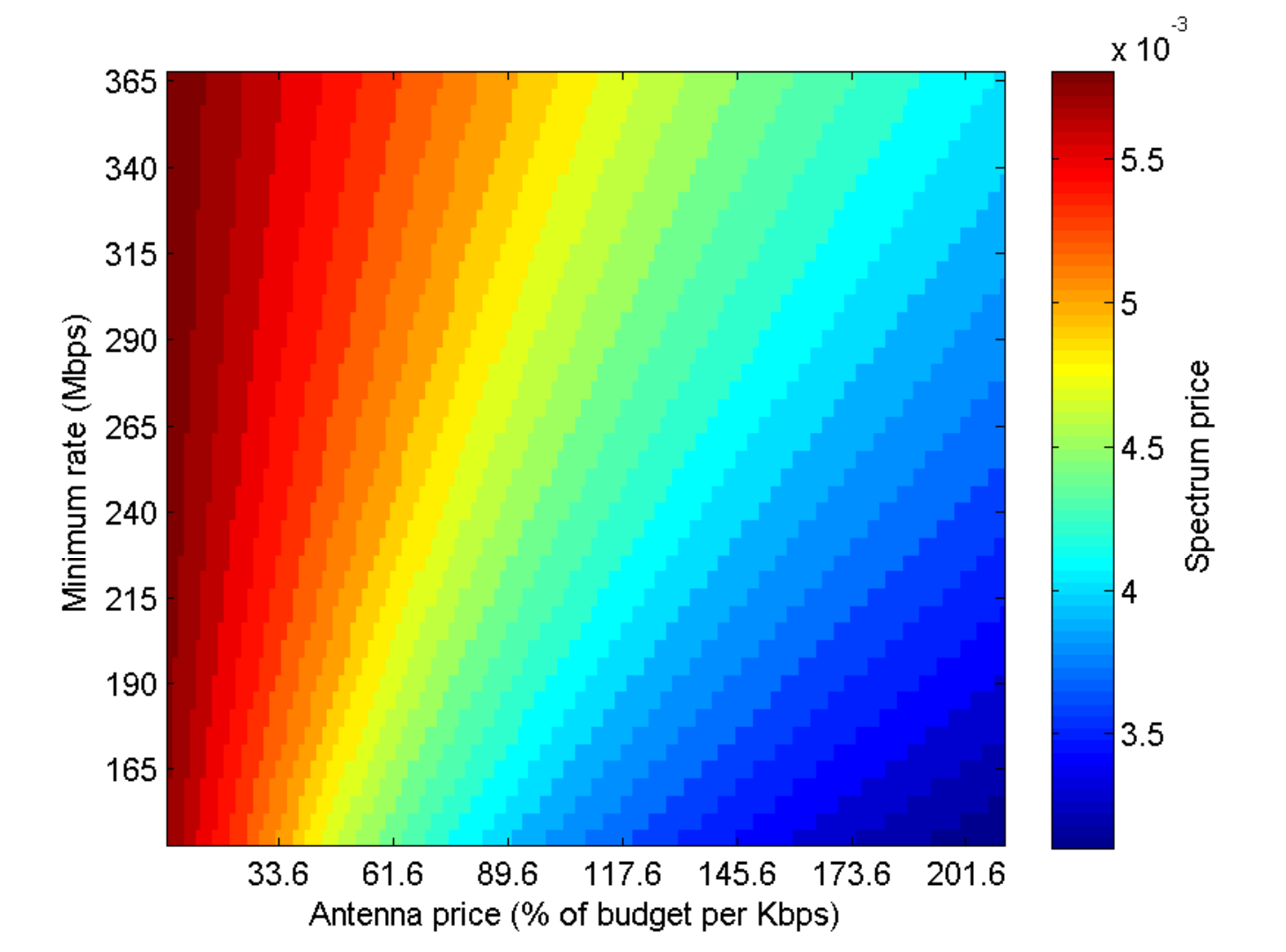}
\caption{\footnotesize Final spectrum price per KHz in correspondence to a particular rate requirement (y axis) and antenna
cost (x axis).}\label{fig:SpectrumCost}
\end{figure}

Figure~\ref{fig:SpectrumCost} shows the final spectrum cost per KHz after the clock phase. In our model increasing the price of antennas causes a reduction in the spectrum price, independently of the required rate. This results from a number of factors: the assumption of a constant maximum budget per Kbps, the requirement on the minimum number of antennas, the complementarity of antennas and spectrum.  When the infrastructure cost increases, bidders opt for acquiring more spectrum. However, due to the complementarity of the two types of resources and, eventually, to the requirement on the minimum number of antennas, a larger fraction of the budget is used for the acquisition of antennas. Hence, the remaining part of the budget, which determines the length of the clock phase and, consequently, the final spectrum price, decreases. This determines the reduction in the spectrum clearing price observed in Figure~\ref{fig:SpectrumCost}. These results show that a lower infrastructure cost does not just imply a better overall revenue for the spectrum provider, but also a higher revenue per unit of auctioned bandwidth.

\section{Discussion} \label{sec:conc}
The complex relation between the different resources that compose a network is bound to have important repercussions on the alternative modes of ownership that are emerging for future wireless networks. In this paper we focused on two of these resources, spectrum and antennas, and their interplay in an auction-based allocation mechanism. 

The model discussed in this paper assumes that antennas can be shared among different operators. Due to its non-rival nature, antennas cannot be sold at auction and their price is fixed. However, to eliminate the potential risk exposure that bidders would face if they acquired spectrum and antennas separately, we devised a combined allocation mechanism so that bidders can obtain the exact bundle of antennas and spectrum that they need. 

Economics theory teaches us that complementary monopolies are typically less efficient than monopolies that hold all the complementary goods. This is because, in the cases of complementary monopolies, both firms benefit from increasing their prices to increase total revenue while simultaneously reducing demand for the complementary good. This results in a reduced overall demand in the market. This is the trend we observe in our results, even though antennas and spectrum are also partially substitute goods. When antennas become more expensive, a smaller number of bidders can be satisfied and a lower combined revenue is obtained. It is possible to counteract this effect by negotiating an agreement between the antenna and the spectrum providers before the auction. For example, the agreement could include a provision under which a proportion of the total revenue is paid to the antenna provider. Antenna access would be at a very low charge that appears as a maintenance access fee. Another possibility is to view the antennas as a platform connecting VNOs to spectrum. From platform economics theory, the platform can decide how to split the usage charge between the two sides of the market. For example the antenna provider could charge the spectrum provider access to the antennas and provide the antennas for free to the VNOs.  

Even more complex scenarios arise when we consider multiple spectrum and antenna providers pooling their resources. The case of multiple spectrum providers is of particular interest in the context of the LSA framework and the PCAST three-tiered spectrum-sharing model, due to the presence of multiple incumbents in both cases. Also, in 2016 FCC will hold the Broadcast Television Incentive Auction in which TV broadcasters will voluntarily bid to relinquish their spectrum usage rights in exchange for a cash payout (reverse auction) and wireless providers will bid to acquire the relinquished spectrum (forward auction). An evolution of the work presented in this paper will be to investigate the impact of acquiring infrastructure and spectrum in these scenarios and to design suitable mechanisms to mitigate the effects of possible divergent interests between the different stakeholders.


\bibliographystyle{IEEEtran}
\bibliography{DMimo_v8}

\end{document}